\documentclass[aps,pra,10pt,showpacs,amsmath,longbibliography,twocolumn,floatfix,eqsecnum]{revtex4-1}
\usepackage[dvipsnames]{xcolor}
\usepackage[colorlinks=true,bookmarks=false,linkcolor=NavyBlue,urlcolor=NavyBlue,citecolor=NavyBlue,breaklinks]{hyperref}
\usepackage{amsmath}
\usepackage{amsfonts}
\usepackage{tabularx}
\usepackage{graphicx}
\usepackage{times}
\usepackage{mathtools}
\usepackage{braket}

\newcommand{\parti}[2]{\frac{\partial #1}{\partial #2}}

\newcommand{\diff}[2]{\frac{d #1}{d #2}}

\newcommand{\avg}[1]{\langle#1\rangle}
\newcommand{\Avg}[1]{\left\langle#1\right\rangle}

\newcommand{\sinc}{\operatorname{sinc}}

\newcommand{\abs}[1]{\left|#1\right|}
\newcommand{\bk}[1]{\left(#1\right)}
\newcommand{\Bk}[1]{\left[#1\right]}
\newcommand{\BK}[1]{\left\{#1\right\}}

\newcommand{\expect}{\mathbb E}
\newcommand{\var}{\mathbb V}

\newcommand{\diag}{\operatorname{diag}}

\newcommand{\cspan}{\operatorname{\overline{span}}}
\newcommand{\nchoosek}[2]
{\begin{pmatrix}#1\\ #2\end{pmatrix}}

\newcommand{\on}[1]{\operatorname{#1}}

\newcommand{\iverson}[1]{1_{#1}}

\DeclareMathOperator*{\argmin}{arg\,min}

\begin{document}

\title{Semiparametric estimation for incoherent optical imaging}

\author{Mankei Tsang}
\email{mankei@nus.edu.sg}
\homepage{https://www.ece.nus.edu.sg/stfpage/tmk/}
\affiliation{Department of Electrical and Computer Engineering,
  National University of Singapore, 4 Engineering Drive 3, Singapore
  117583}

\affiliation{Department of Physics, National University of Singapore,
  2 Science Drive 3, Singapore 117551}

\date{\today}

%\pacs{42.50.Wk, 03.65.Ta, 42.65.Yj}

\begin{abstract}
  The theory of semiparametric estimation offers an elegant way of
  computing the Cram\'er-Rao bound for a parameter of interest in the
  midst of infinitely many nuisance parameters.  Here I apply the
  theory to the problem of moment estimation for incoherent imaging
  under the effects of diffraction and photon shot noise. Using a
  Hilbert-space formalism designed for Poisson processes, I derive
  exact semiparametric Cram\'er-Rao bounds and efficient estimators
  for both direct imaging and a quantum-inspired measurement method
  called spatial-mode demultiplexing (SPADE).  The results establish
  the superiority of SPADE even when little prior information about
  the object is available.
\end{abstract}

\maketitle
\section{Introduction}
Two fundamental problems confront incoherent optical imaging: the
diffraction limit \cite{born_wolf,goodman} and the photon shot noise
\cite{mandel,goodman_stat}. To quantify their effects on the
resolution rigorously, the Cram\'er-Rao bound (CRB) on the error of
parameter estimation \cite{lehmann98} has been widely used, especially
in astronomy and fluorescence microscopy
\cite{farrell66,tsai79,zmuidzinas03,feigelson,ram,small14,deschout,chao16,diezmann17,bettens,vanaert,villiers}. Most
previous studies, however, assume that the object has a simple
specific shape, such as a point source or two, and only one or few
parameters of the object are unknown. Such parametric models may not
be justifiable when there is little prior information about the
object. Without a parametric model, the CRB seems
intractable---infinitely many parameters are needed to specify the
object distribution, leading to a Fisher information matrix with
infinitely many entries, and then the infinite-dimensional matrix has
to be inverted to give the CRB.  While there also exist many studies
on superresolution that can deal with more general objects
\cite{villiers,candes13,candes14,schiebinger}, they either ignore
noise or use noise models that are too simplistic to capture the
signal-dependent nature of photon shot noise.

To compute the CRB and to evaluate the efficiency of estimators for
general objects, here I propose a theory of semiparametric estimation
for incoherent optical imaging.  Semiparametric estimation refers to
the estimation of a parameter of interest in the presence of
infinitely many other unknown ``nuisance'' parameters
\cite{bickel93,tsiatis06}. The method has found many applications in
econometrics, biostatistics, and astrostatistics \cite{bickel93}. A
typical example is the estimation of the mean of a random variable
when its probability density is assumed to have finite variance but
otherwise arbitrary. Thanks to a beautiful Hilbert-space formalism
\cite{bickel93,tsiatis06}, the semiparametric theory is able to
compute the CRB for such problems despite the infinite dimensionality
and also evaluate the existence and efficiency of semiparametric
estimators. Such problems are exactly the type that bedevil the study
of imaging thus far, and here I show how the semiparametric theory can
be used to yield similarly elegant results for optical imaging.

The optics problem of interest here is the far-field imaging of an
object emitting spatially incoherent light
\cite{goodman,goodman_stat}, with the most important applications
being optical astronomy \cite{farrell66,tsai79,zmuidzinas03,feigelson}
and fluoresence microscopy
\cite{ram,small14,deschout,chao16,diezmann17}.  With a finite
numerical aperture, the imaging system introduces a spatial bandwidth
limit to the waves, otherwise known as the diffraction limit
\cite{born_wolf,goodman}. The standard measurement, called direct
imaging, records the intensity of the light on the image
plane. Recently, quantum information theory inspired the invention of
an alternative measurement called spatial-mode demultiplexing (SPADE)
\cite{tnl}, which has been shown theoretically
\cite{tnl,tsang17,tsang18a,tsang19,zhou19,dutton19,sliver,tnl2,nair_tsang16,lupo,tsang18,ant,lu18,rehacek17,yang17,kerviche17,chrostowski17,rehacek17a,rehacek18,backlund18,napoli19,yu18,prasad19,larson18,tsang_comment19,larson19,grace19,bonsma19,tsang19a}
and experimentally
\cite{tang16,tham17,paur16,yang16,donohue18,hassett18,zhou19a} to be
superior to direct imaging in resolving two sub-Rayleigh sources and
estimating the size and moments of a subdiffraction object. Most of
the aforementioned studies, however, assume parametric models for the
object.  Exceptions include
Refs.~\cite{yang16,tsang17,tsang18a,tsang19,zhou19,bonsma19}, which
consider the estimation of object moments, but the results there are
not conclusive---only the CRB for direct imaging was computed exactly
\cite{tsang18a}, while the CRB for SPADE was evaluated only
approximately \cite{tsang17,tsang18a,zhou19}. Another problem is the
existence and efficiency of unbiased moment estimators; again only
approximate results have been obtained so far
\cite{tsang17,tsang18a}. Building on the established semiparametric
theory \cite{bickel93,tsiatis06}, here I compute the exact
semiparametric CRBs and also propose unbiased and efficient moment
estimators for both direct imaging and SPADE. These results enable a
fair and rigorous comparison of the two measurement methods, which
proves the fundamental superiority of SPADE for moment estimation.

This paper is organized as follows. Section~\ref{sec_crb} introduces
the Fisher information and the CRB for Poisson processes.
Section~\ref{sec_hilbert} presents the semiparametric CRB in terms of
a Hilbert-space formalism designed for such
processes. Section~\ref{sec_imaging} introduces the models of direct
imaging and SPADE.  Section~\ref{sec_direct} computes the CRB for
moment estimation with direct imaging and proposes an efficient
estimator.  Section~\ref{sec_ccrb} shows how the CRB should be
modified for a normalized object distribution.
Section~\ref{sec_spade} computes the CRB for SPADE and also proposes
an efficient estimator. Section~\ref{sec_compare} uses the CRBs to
compare the performances of direct imaging and SPADE, demonstrating
the superiority of SPADE for subdiffraction
objects. Section~\ref{sec_conc} concludes the paper and points out
open issues, while the Appendices detail the technical issues that
arise in the main text.

\section{\label{sec_crb}Cram\'er-Rao bound for Poisson processes}
For optical astronomy
\cite{farrell66,goodman_stat,zmuidzinas03,feigelson}, fluorescence
microscopy \cite{ram,small14,deschout,chao16,diezmann17}, and even
electron microscopy \cite{bettens,vanaert}, Poisson noise can be
safely assumed. Suppose that each detector in a photodetector array is
labeled by $x \in \mathcal X$, where $\mathcal X$ denotes the detector
space.  Assume that the observed process, such as the image recorded
by a camera, is a Poisson random measure $n$ on $\mathcal X$ and its
$\sigma$-algebra $\Sigma$, with a mean given by the intensity measure
$\bar n$ on the same $(\mathcal X,\Sigma)$ \cite{cinlar}.
$n(\mathcal A)$ for any $\mathcal A \in \Sigma$ is then a Poisson
variable with mean $\bar n(\mathcal A)$. For example, if
$\mathcal X \subseteq \mathbb R^2$ is a two-dimensional surface, then
$\Sigma$ is the set of all subareas that can be defined on the
surface, and $n(\mathcal A) = \int_{x \in \mathcal A} dn(x)$ is the
detected photon number over the area $\mathcal A$.  For any vectoral
function $h:\mathcal X \to \mathbb R^{q}$ on the detector space,
\begin{align}
\check h(n) &= \int h(x) dn(x),
\end{align}
a linear functional of $n$, is a random variable with statistics
\begin{align}
  \expect(\check h) &= \int h(x) d\bar n(x) = \nu(h),
\label{poisson_mean}
  \\
  \var(\check h) &= \expect(\check h\check h^\top) 
- \expect(\check h)\expect(\check h^\top)
  = \nu(hh^\top),
\label{poisson_covar}
\end{align}
where $\expect$ denotes the statistical expectation, $\var$ denotes
the covariance, $\nu$ denotes the average with respect to the
intensity measure $\bar n$, $\top$ denotes the matrix transpose, and
all vectors in this paper are column vectors.

Suppose that $\bar n$ depends on an unknown vectoral parameter
$\theta \in \Theta \subset \mathbb R^{p}$ with $p$ entries and has a
density $f(x|\theta)$ with respect to a dominating measure $\mu$ such
that $f(x|\theta) = d\bar n(x|\theta)/d\mu(x)$. The log-likelihood
derivatives are given by \cite{snyder_miller}
\begin{align}
\check S_j(n|\theta) &=  
\int \parti{}{\theta_j}\ln f(x|\theta) dn(x)
-\parti{}{\theta_j}\int d\bar n(x|\theta).
\label{ld}
\end{align}
As $\check S$ is a linear functional of $n$, its covariance, called
the Fisher information matrix, can be simplified via
Eq.~(\ref{poisson_covar}) and is given by \cite{snyder_miller}
\begin{align}
J &= \var(\check S) =\int S(x) [S(x)]^\top d\bar n(x)  = 
\nu(SS^\top),
\end{align}
where $S$ is a vector of detector-space functions given by
\begin{align}
  S_j(x|\theta) &= \parti{}{\theta_j} \ln f(x|\theta).
\end{align}
Here $\var$, $\check S$, $\bar n$, $S$, and $\nu$ are all evaluated at
the same $\theta$, and I assume hereafter that all functions of
$\theta$ are evaluated implicitly at the same $\theta$. Each $S_j$ is
hereafter called a score function, borrowing the same terminology for
$\check S$ in statistics \cite{bickel93,tsiatis06}. An important
distinction is that, whereas $\expect(\check S_j) = 0$, $\nu(S_j)$
does have to be zero, since $\bar n$ does not have to be normalized.

Let $\beta(\theta)$ be a scalar parameter of interest.  If
$\beta(\theta) = \theta_k$ for example, then all the other parameters
in $\theta$ are called nuisance parameters. For any unbiased estimator
$\check\beta(n)$, the CRB on its variance is \cite{lehmann98}
\begin{align}
\var(\check\beta) &\ge
u^\top J^{-1} u = {\rm CRB},
&
u_j &= \parti{\beta}{\theta_j}.
\label{crb}
\end{align}
$J^{-1}$ seems intractable if $\theta$ is infinite-dimensional.  The
next section introduces a cleverer method.

\section{\label{sec_hilbert}Semiparametric Cram\'er-Rao bound}

\begin{figure}[htbp!]
\centerline{\includegraphics[width=0.48\textwidth]{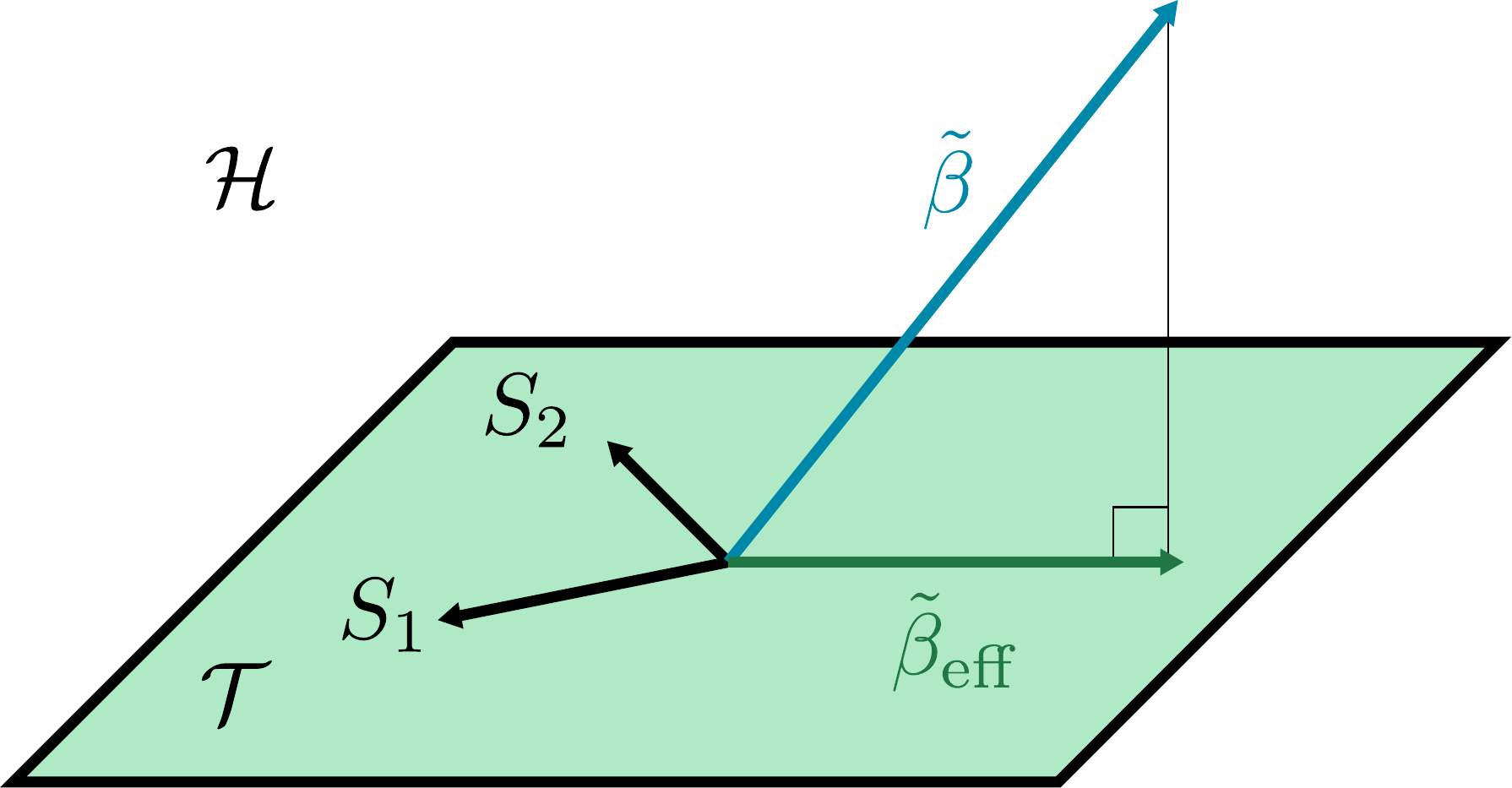}}
\caption{\label{efficient_influence}The efficient influence
  $\tilde\beta_{\rm eff}$ is the orthogonal projection of any
  influence function $\tilde\beta \in \mathcal H$ that satisfies
  Eq.~(\ref{delta}) into the tangent space
  $\mathcal T = \on{span}(S)$.  The norm of $\tilde\beta_{\rm eff}$
  gives the CRB.}
\end{figure}

The key to the semiparametric theory is to treat random variables as
elements in a Hilbert space \cite{bickel93,tsiatis06}. Here I
introduce another Hilbert space for detector-space functions on top of
the statistical one for the purpose of computing the CRB for Poisson
processes.  Define an inner product between two scalar functions
$h_1,h_2:\mathcal X \to \mathbb R$ as
\begin{align}
\Avg{h_1,h_2} &= \nu(h_1h_2) = 
\int h_1(x)h_2(x)d\bar n(x),
\end{align}
and the norm as
\begin{align}
||h|| &= \sqrt{\Avg{h,h}} = \sqrt{\nu(h^2)}.
\end{align}
With the inner product, a Hilbert space $\mathcal H$ can be defined as
the set of all square-summable functions, viz.,
\begin{align}
\mathcal H &= \BK{h(x): \nu(h^2) < \infty}.
\label{hilbert}
\end{align}
Denote the set of score functions $\{S_j\}$ as $S$ in a slight abuse
of notation. If the Fisher information $J_{jj} = \nu(S_j^2) < \infty$
for all $j$, $S \subset \mathcal H$. Define the tangent space
$\mathcal T \subseteq \mathcal H$ of a parametric model as the linear
span of $S$, or
\begin{align}
\mathcal T &= \BK{w^\top S: w \in \mathbb R^{p}}
=\on{span}(S).
\label{tangent}
\end{align}
Define also an ``influence'' function as any
$\tilde\beta \in \mathcal H$ that satisfies
\begin{align}
\nu(\tilde\beta S) &= u,
\label{delta}
\end{align}
borrowing the name of a similar concept in statistics
\cite{bickel93,tsiatis06}. The Cauchy-Schwartz inequality
$\nu(\tilde\beta^2) [w^\top\nu(SS^\top) w] \ge (u^\top w)^2$ with
$w = [\nu(SS^\top)]^{-1} u$ then yields
\begin{align}
\nu(\tilde\beta^2) &\ge u^\top J^{-1} u,
\label{schwartz}
\end{align}
the right-hand side of which coincides with the CRB given by
Eq.~(\ref{crb}). Define the efficient influence as the influence
function that saturates Eq.~(\ref{schwartz}), viz.,
\begin{align}
\tilde\beta_{\rm eff} &= u^\top J^{-1} S = \nu(\tilde\beta S^\top)
\Bk{\nu(SS^\top)}^{-1} S,
\label{beta_eff}
\\
{\rm CRB} &= \nu(\tilde\beta_{\rm eff}^2).
\label{CRB}
\end{align}
Equation~(\ref{beta_eff}) can be interpreted as the orthogonal
projection of any influence function $\tilde\beta \in \mathcal H$ that
satisfies Eq.~(\ref{delta}) into $\mathcal T$, viz.,
\begin{align}
  \tilde\beta_{\rm eff} &= \Pi(\tilde\beta|\mathcal T) 
= \argmin_{h \in \mathcal T}||\tilde\beta - h||.
\label{influ_eff}
\end{align}
Figure~\ref{efficient_influence} illustrates this concept. 

Consider now the semiparametric scenario. For the purpose of this
paper, it suffices to assume that the dimension of $\theta$ is
infinite but countable ($p = \infty$). The score functions are still
defined in the same way, but now there are infinitely many of
them. The tangent space should be modified to be the closed linear
span
\begin{align}
  \mathcal T &= \cspan(S),
\label{tangent2}
\end{align}
so that projection into it is well defined \cite{reed_simon}, and the
semiparametric CRB is still given by Eqs.~(\ref{delta}), (\ref{CRB}),
and (\ref{influ_eff}); see Appendix~\ref{sec_proof} for a proof. This
Hilbert-space approach is tractable when finding a candidate influence
function is straightforward and the tangent space is so large that the
candidate is already in it or at least very close to it.  If the
dimension of $\theta$ is uncountable, the tangent space and the CRB
can still be defined via the concept of parametric submodels
\cite{bickel93,tsiatis06}, although it is not needed here.

If $\beta$ can be expressed as a functional $\beta(f)$, a useful way
of finding an influence function is to consider a functional
derivative of $\beta(f)$ with respect to $h(x)$ defined as
\begin{align}
\dot\beta(f,h)
&= \lim_{\epsilon \to 0} 
\frac{\beta((1+ \epsilon h)f)-\beta(f)}{\epsilon}
\\
&= \int \tilde\beta(x) h(x) f(x)  d\mu(x)  = \nu(\tilde\beta h),
\end{align}
which leads to
\begin{align}
\parti{\beta}{\theta_j}
&= \lim_{\epsilon \to 0} \frac{\beta(f + \epsilon \partial f/\partial \theta_j)
-\beta(f)}{\epsilon}
\\
&= \dot\beta(f,S_j)= \nu(\tilde\beta S_j)= u_j,
\end{align}
and the $\tilde\beta(x)$ function obtained from the functional
derivative is an influence function that satisfies
Eq.~(\ref{delta}). The simplest example is the linear functional
\begin{align}
\beta(f) &= \int \tilde\beta(x)f(x) d\mu(x) = \nu(\tilde\beta),
\label{linear_functional}
\end{align}
and $\tilde\beta(x)$ is an influence function. 

If the tangent space is so large that $\mathcal T = \mathcal H$, then
a square-summable influence function is already in
$\mathcal H = \mathcal T$ and therefore efficient.  There are often
some constraints that make $\mathcal T$ smaller, however, and the CRB
is reduced as a result.  For example, if the constraint can be
expressed as
\begin{align}
\gamma(f) &= 0,
\end{align}
and its functional derivative is
\begin{align}
\dot\gamma(f,h) &= \nu(h\tilde\gamma),
\end{align}
then
\begin{align}
\parti{\gamma(f)}{\theta_j} &= 
\dot\gamma(f,S_j) = \nu(\tilde\gamma S_j) = \Avg{\tilde\gamma,S_j} = 0,
\end{align}
and it follows that $\tilde\gamma$ should be placed in the set that
spans $\mathcal T^\perp$, the orthocomplement of $\mathcal T$ in
$\mathcal H$.  In terms of $\mathcal T^\perp$, the efficient influence
can be evaluated as
\begin{align}
  \tilde\beta_{\rm eff} &= \tilde\beta-\Pi(\tilde\beta|\mathcal T^\perp).
\label{newbeta}
\end{align}
If $\mathcal T^\perp = \on{span}(R)$, then
\begin{align}
\Pi(\tilde\beta|\mathcal T^\perp) &= \nu(\tilde\beta R^\top)
\Bk{\nu(RR^\top)}^{-1} R,
\label{project_ortho}
\end{align}
which is still tractable if $R$ has a low dimension.

\section{\label{sec_imaging}Incoherent optical imaging}
\begin{figure}[htbp!]
\centerline{\includegraphics[width=0.48\textwidth]{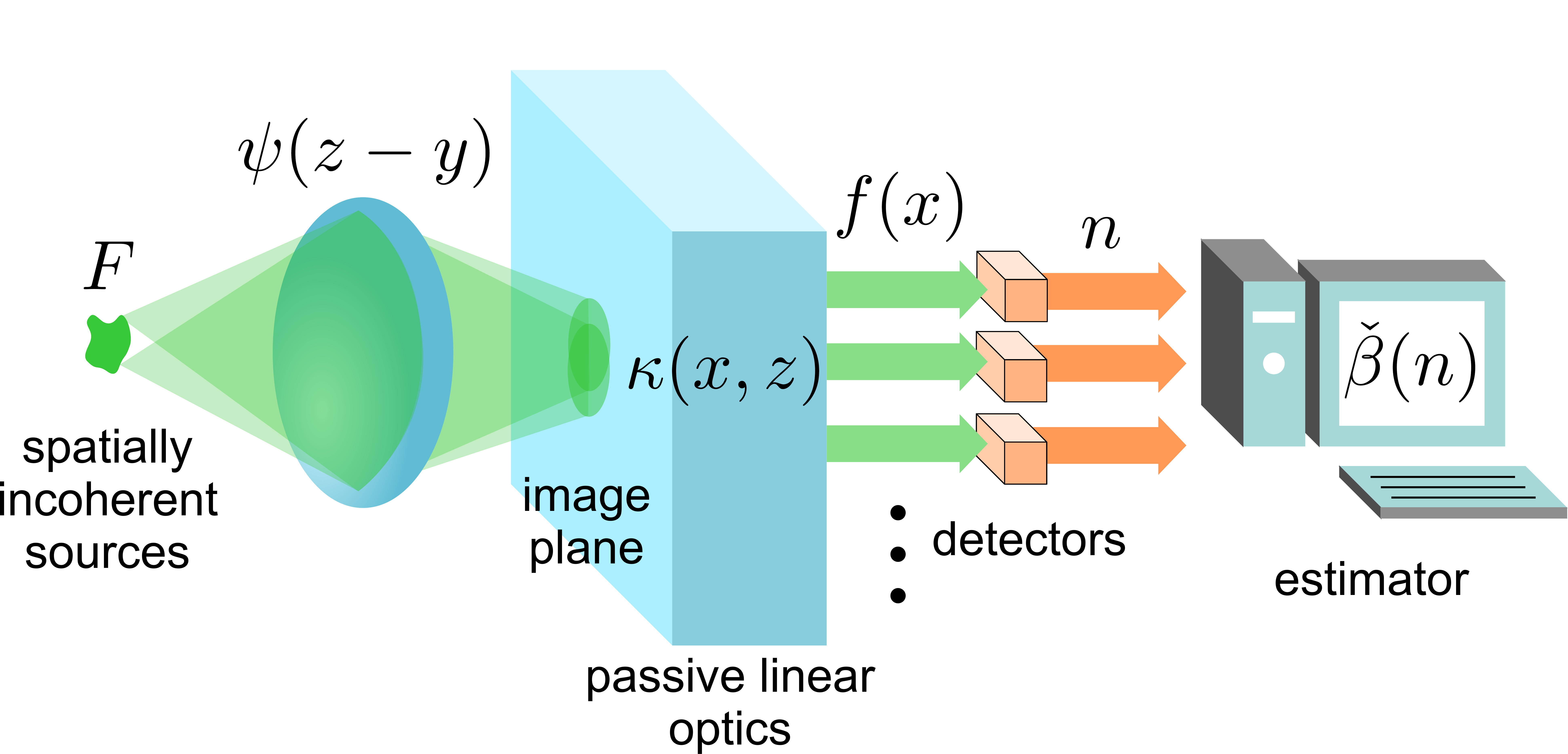}}
\caption{\label{fig_spade}A far-field incoherent imaging system. See
  the main text for definitions.}
\end{figure}

Consider a distribution of spatially incoherent sources described by
the measure $F$ on the object plane with coordinate $y$, a far-field
paraxial imaging system with point-spread function $\psi(z-y)$ for the
field \cite{goodman}, further processing of the field on the image
plane with coordinate $z$ via passive linear optics with Green's
function $\kappa(x,z)$, and Poisson noise at the output detectors
labeled by $x \in \mathcal X$, as depicted by
Fig.~\ref{fig_spade}. For simplicity, assume one-dimensional imaging
such that $y, z \in \mathbb R$; generalization for two-dimensional
imaging is possible \cite{tsang17,tsang18a} but not very
interesting. The intensity can be described by the mixture model
\cite{tnl,tnl2,tsang17,tsang18a,goodman_stat}
\begin{align}
  f(x) &= 
\int \abs{\int \kappa(x,z)\psi(z-y)dz}^2dF(y),
\label{imaging}
\end{align}
where the image-plane coordinate $z$ is normalized with respect to the
magnification factor, both $y$ and $z$ are normalized with respect to
the width of the point-spread function such that they are
dimensionless, and $\psi$ is normalized as $\int |\psi(x)|^2dx = 1$.
This semiclassical Poisson model can be derived from standard quantum
optics \cite{stellar,tnl,tsang19a}.

For direct imaging with infinitesimal pixels,
$\kappa(x,z) = \sqrt{\tau}\delta(x-z)$, where $\tau$ is a positive
conversion factor, $x \in \mathcal X = \mathbb R$ denotes the position
of each pixel, $d\mu(x) = dx$, and the image intensity obeys the
convolution model
\begin{align}
f(x) &= \int H(x-y)dF(y),
&
H(x) &= \tau |\psi(x)|^2,
\label{direct}
\end{align}
which will be studied in Sec.~\ref{sec_direct}.

The most remarkable physics of the problem lies in the possibility of
improving the measurement via optics with a different Green's function
$\kappa$.  Quantum information theory has shown that substantial
improvement is possible for subdiffraction objects, and SPADE has been
found to be quantum-optimal in many special cases
\cite{tnl,tsang17,tsang18a,tsang19,zhou19,dutton19,sliver,tnl2,nair_tsang16,lupo,tsang18,ant,lu18,rehacek17,yang17,kerviche17,chrostowski17,rehacek17a,rehacek18,backlund18,napoli19,yu18,prasad19,larson18,tsang_comment19,larson19,grace19,bonsma19,tsang19a}. In
one version of SPADE, $\kappa^*(q,z)$ is the $q$th mode function in
the point-spread-function-adapted (PAD) basis
\cite{rehacek17,tsang18a}, such that the output intensity is given by
\begin{align}
f(q) &= \int H(q|y) dF(y),
\quad
q \in \mathcal X = \mathbb N_0,
\label{spade}
\\
H(q|y) &= \abs{\int \kappa(q,z)\psi(z-y)dz}^2,
\label{Hq}
\end{align}
where $\mu$ is simply the counting measure and $\kappa$ and $H$ obey
special properties, as further discussed in Sec.~\ref{sec_spade}.  For
a fair comparison, the quantum efficiencies of direct imaging and
SPADE are assumed to be the same, meaning that \cite{tsang18a}
\begin{align}
\sum_{q=0}^\infty H(q|y) &= \tau,
\label{conserve}
\end{align}
where $\tau$ is the same factor as that for direct imaging.  Then
\begin{align}
N &= \expect[n(\mathcal X)] = \bar n(\mathcal X) = \nu(1) = \tau \int dF(y),
\label{N}
\end{align}
the expected photon number received in total, is also the same.

\section{\label{sec_direct}Moment estimation
with direct imaging}
Consider the direct-imaging model given by Eq.~(\ref{direct}).  Assume
that $H$ can be expanded in a Taylor series as
\begin{align}
H(x-y) &= \sum_{j=0}^\infty \frac{(-1)^j}{j!}
\parti{^j H(x)}{x^j} y^j,
\end{align}
which leads to
\begin{align}
f(x) &=  \sum_{j=0}^\infty \frac{(-1)^j}{j!}
\parti{^j H(x)}{x^j}\theta_j,
\label{parametric}
\end{align}
where the unknown parameters are the object moments defined by
\begin{align}
  \theta_j &= \int   y^j dF(y),\quad j \in \mathbb N_0.
\label{moments}
\end{align}
For the CRB to hold, the parameter space should correspond to the
condition that $F$ contains an infinite number of point sources with
different positions, as discussed in
Appendix~\ref{sec_moments}. Appendix~\ref{sec_series} shows a way of
reconstructing $F$ from $\theta$ via an orthogonal-series expansion,
following Ref.~\cite{tsang19a}.

The score function for each $\theta_j$ is
\begin{align}
S_j(x) &= \frac{(-1)^j}{j!f(x)}\parti{^j H(x)}{x^j}.
\label{S_direct}
\end{align}
It turns out that the tangent space $\mathcal T$ for this problem is
equal to the whole Hilbert space $\mathcal H$ under certain technical
conditions, as shown in Appendix~\ref{sec_tangent}.

Let the parameter of interest be
\begin{align}
\beta &= u^\top\theta = \sum_{j=0}^\infty u_j\theta_j,
\label{beta}
\end{align}
where $u$ is independent of $\theta$. To find a candidate influence
function, a trick \cite{meister06} is to consider the image moments
$\phi$ given by
\begin{align}
\phi &= \int \tilde\phi(x) d\bar n(x) = \nu(\tilde\phi),
\label{phi}
\end{align}
where
\begin{align}
\tilde\phi_j(x) &= x^j,
\quad
j \in \mathbb N_0
\label{monomials}
\end{align}
are the monomials.  Assuming that all the moments of $F$ and $H$ are
finite such that all the moments of $f$ are also finite, $\phi$ can be
related to the object moments via
\begin{align}
\phi_j &=  \iint x^j H(x-y)dF(y) dx
\\
&=  \iint  (z+y)^j H(z) dF(y) dz
\\
&=  \iint \sum_{k=0}^j \nchoosek{j}{k} z^{j-k}y^k H(z) dF(y) dz
\\
&=  \sum_{k=0}^\infty C_{jk}\theta_k,
\end{align}
where
\begin{align}
C_{jk} &= \iverson{j\ge k}\nchoosek{j}{k}\int  H(x) x^{j-k} dx
\label{C_direct}
\end{align}
and
\begin{align}
\iverson{\rm proposition} &= 
\begin{cases}
  1 & \textrm{if proposition is true},\\
  0 & \textrm{otherwise}.
\end{cases}
\end{align}
$C$ is a lower-triangular matrix, and with
$C_{jj} =\int H(x)dx =\tau > 0$, $C^{-1}$ is well defined and also
lower-triangular even if the dimension of $C$ is infinite, as shown in
Appendix~\ref{sec_triangular}. The object moments can then be related
to the image moments by
\begin{align}
\theta &= C^{-1}\phi,
\end{align}
and $\beta$ can be expressed as
\begin{align}
\beta = u^\top\theta = u^\top C^{-1}\phi = \nu(u^\top C^{-1}\tilde\phi).
\end{align}
According to Eq.~(\ref{linear_functional}), an influence function is
\begin{align}
\tilde\beta(x) &= 
u^\top C^{-1}\tilde\phi(x) = u^\top \tilde\theta(x),
\label{influ_direct}
\\
\tilde\theta(x) &= C^{-1} \tilde\phi(x).
\label{tilde_theta}
\end{align}
Since $\mathcal T = \mathcal H$ as shown in
Appendix~\ref{sec_tangent}, the $\tilde\beta$ given by
Eq.~(\ref{influ_direct}) belongs to $\mathcal H = \mathcal T$ and is
efficient according to Eq.~(\ref{influ_eff}) as long as it is
square-summable. For example, if $u$ contains a finite number of
nonzero entries, $\tilde\beta$ is a polynomial of $x$ and must be
square-summable, since all the moments of $f$ are assumed to be
finite.  The CRB is hence
\begin{align}
{\rm CRB}^{(\rm direct)} 
&= \nu(\tilde\beta^2) = u^\top \nu(\tilde\theta\tilde\theta^\top) u
\nonumber\\
&= u^\top C^{-1}\nu(\tilde\phi\tilde\phi^\top)  C^{-\top}u,
\label{CRB_direct}
\end{align}
where $C^{-\top} = (C^{-1})^\top$.  This result coincides with that
derived in Ref.~\cite{tsang18a} via a more direct but less rigorous
method, which is repeated in Appendix~\ref{sec_direct2} for
completeness.

An unbiased and efficient estimator $\check\beta(n)$ in terms of the
observed process $n$ can be constructed from the efficient influence
as
\begin{align}
\check\beta(n) &= \int \tilde\beta(x) dn(x) = u^\top\check\theta(n),
\label{est_direct}
\end{align}
where the object moment estimator is
\begin{align}
\check\theta(n) &= C^{-1}\check\phi(n),
&
\check\phi(n) &= \int \tilde\phi(x) dn(x).
\end{align}
$\check\beta(n)$ is a linear filter of $n$, so its variance is
$\var(\check\beta) = \nu(\tilde\beta^2)$, which coincides with the
CRB. It is important to note that this estimator does not require any
knowledge of the unknown parameters, as $\check\phi(n)$ is simply the
empirical moments of the observed image, and $C^{-1}$ is a
lower-triangular matrix that depends on the moments of the
point-spread function $H$. The estimator still works even if the
object happens to consist of a finite number of point sources and
$\theta$ is on the boundary of the parameter space, although its
efficiency in that case is a more difficult question, as explained in
Appendix~\ref{sec_moments}.  Unlike some of the previous studies on
superresolution \cite{villiers,candes13,candes14,schiebinger}, the
results here place no restriction on the separations of the point
sources and also account for Poisson noise explicitly.

\section{\label{sec_ccrb}Constrained Cram\'er-Rao bound}
In imaging, the parameters of interest are often the moments with
respect to a normalized object distribution with $\int dF(y) = 1$.  A
simple way of modeling this is to assume that $\theta_0 = 1$ is
known. This constraint also makes the model comparable to those in
Refs.~\cite{tsang17,tsang19,tsang19a}.  Then
\begin{align}
  N &= \phi_0 = \tau\theta_0
\end{align}
is known as well, implying the constraint
$\gamma(f) = \int f(x)dx - N= 0$. The constraint can be differentiated
to yield $\dot\gamma(f,S_j) = \nu(S_j) = \avg{S_j,1} = 0$, leading to
$\mathcal T^\perp = \on{span}(1)$.  The new efficient influence,
according to Eqs.~(\ref{newbeta}) and (\ref{project_ortho}), should
therefore be
\begin{align}
\tilde\beta_{\rm eff} &= \tilde\beta-\Pi(\tilde\beta|\mathcal T^\perp)
= \tilde\beta - \frac{\nu(\tilde\beta)}{\nu(1)} = \tilde\beta
- \frac{\beta}{N}.
\end{align}
The constrained CRB is now
\begin{align}
{\rm CRB}_{\theta_0 = 1}^{({\rm direct})} &= \nu(\tilde\beta_{\rm eff}^2)  
= \frac{1}{N}\Bk{\nu_0(\tilde\beta_0^2)-\beta^2},
\label{CCRB}
\\
\tilde\beta_0(x) &=  N\tilde\beta(x) = u^\top (C/\tau)^{-1}\tilde\phi(x),
\end{align}
where $\nu_0(h) = \nu(h)/\nu(1)$ is the normalized version of
$\nu$. The CRB is necessarily lowered by the constraint.  Other
approaches to the constrained CRB yield the same result, as discussed
in Appendix~\ref{sec_ccrb2}.

To construct a near-efficient estimator, suppose that
$n(\mathcal X) = \int dn(x) = L > 0$ photons have been detected.  Then
$dn(x) = \sum_{l=1}^L \iverson{x=X_l}$, and the photon positions
$\{X_1,X_2,\dots,X_L\}$ are independent and identically distributed
according to the probability measure $\bar n/N$. Consider the
estimator
\begin{align}
\check\beta(n) &= 
\frac{1}{L} \int \tilde\beta_0(x) dn(x) = \frac{1}{L}\sum_{l=1}^L \tilde\beta_0(X_l).
\end{align}
It is straightforward to show that
\begin{align}
\expect(\check\beta|n(\mathcal X)=L) &= \nu_0(\tilde\beta_0) = \beta,
\\
\var(\check\beta|n(\mathcal X) = L)&= 
\frac{1}{L}\Bk{\nu_0(\tilde\beta_0^2)-\beta^2},
\end{align}
which is close to the CRB given by Eq.~(\ref{CCRB}) if $L$ is close to
its expected value $N$.  The results are then consistent with standard
results in semiparametric estimation concerning the moments of a
normalized probability measure \cite{bickel93}.

\section{\label{sec_spade}Even-moment estimation 
with SPADE}
Now consider the SPADE model given by Eqs.~(\ref{spade}) and
(\ref{Hq}) and the Fourier transforms
\begin{align}
\Psi(k) &= \frac{1}{\sqrt{2\pi}}\int \psi(z)\exp(-ikz)dz,
\\
\Phi_q(k) &= \frac{1}{\sqrt{2\pi}}\int \kappa^*(q,z)\exp(-ikz) dz.
\end{align}
Suppose that $\Phi = \{\Phi_q(k)\}$ is the PAD basis
\cite{rehacek17,tsang18a} given by
\begin{align}
\Phi_q(k) &= \sqrt{\tau} b_q(k)\Psi(k),
\quad
q \in \mathbb N_0,
\end{align}
where $b = \{b_q(k): q \in \mathbb N_0\}$ is the set of orthonormal
polynomials defined by
\begin{align}
\int  \abs{\Psi(k)}^2 b_q(k) b_r(k) dk = \delta_{qr}.
\label{bq_ortho}
\end{align}
The polynomials exist for all $q \in \mathbb N_0$ as long as the
support of $|\Psi(k)|^2$ is infinite \cite{dunkl}, and the
orthonormality of $\Phi$ ensures that the measurement can be
implemented by passive linear optics
\cite{tnl,tnl2,rehacek17}. Equation~(\ref{Hq}) becomes
\begin{align}
H(q|y) &= \tau \abs{\int  \abs{\Psi(k)}^2 b_q(k)\exp(-iky)dk}^2
\label{Hqy}
\\
&= 
\tau \abs{
\int \abs{\Psi(k)}^2 b_q(k)\sum_{j=0}^\infty \frac{(-iky)^j}{j!} dk}^2.
\end{align}
As the $b$ polynomials are derived by applying the Gram-Schmidt
procedure to the monomials $(1,k,k^2,\dots)^\top$, their basic
properties include $\int |\Psi(k)|^2 b_q(k) k^r dk= 0$ if $r < q$,
$\int |\Psi(k)|^2 b_q(k) k^q dk \neq 0$, and $b_q(k) = (-1)^q b_q(-k)$
if $|\Psi(k)|^2$ is even, as is often the case in optics.  These
properties lead to
\begin{align}
H(q|y) &= \sum_{j=0}^\infty C_{qj} y^{2j},
\label{C_spade}
\end{align}
where $C$ is an upper-triangular matrix ($C_{qj} = 0$ if $j < q$) with
positive diagonal entries ($C_{qq} > 0$). Equation~(\ref{spade}) becomes
\begin{align}
f(q) &= \sum_{j=0}^\infty C_{qj}\theta_{2j},
\label{f_spade}
\end{align}
which depends on the even moments
\begin{align}
\theta_{2j} &= \int y^{2j} dF(y),
\quad
j \in \mathbb N_0.
\label{even_moments}
\end{align}
The score function with respect to each $\theta_{2j}$ becomes
\begin{align}
S_{j}(q) &= \frac{1}{f(q)}
\parti{f(q)}{\theta_{2j}} = \frac{C_{qj}}{f(q)}.
\label{S_spade}
\end{align}
Appendix~\ref{sec_tangent2} proves that
$\mathcal T = \cspan(S) = \mathcal H$.

To find a candidate influence function, suppose that
Eq.~(\ref{f_spade}) can be inverted to give
\begin{align}
\theta_{2j} &= \sum_{q=0}^\infty (C^{-1})_{jq} f(q).
\end{align}
An influence function for $\beta = u^\top\theta$ according to
Eq.~(\ref{linear_functional}) is therefore
\begin{align}
\tilde\beta(q) &= u^\top\tilde\theta(q),
&
  \tilde\theta_{2j}(q) &= (C^{-1})_{jq}.
\label{influ_spade}
\end{align}
Since $\mathcal T = \mathcal H$, this $\tilde\beta$ belongs to
$\mathcal T$ and is efficient as long as it is square-summable. The
CRB is hence
\begin{align}
{\rm CRB}^{(\rm SPADE)} &= \nu(\tilde\beta^2) = 
u^\top\nu(\tilde\theta\tilde\theta^\top)u
\nonumber\\
&= u^\top C^{-1}D C^{-\top}u,
\label{CRB_spade}
\\
D_{jk} &= f(j)\delta_{jk}.
\label{D}
\end{align}
A more direct but heuristic way of deriving Eq.~(\ref{CRB_spade}) is
shown in Appendix~\ref{sec_spade2}. An unbiased and efficient
estimator in terms of the observed detector counts $n$ is
\begin{align}
\check\beta(n) &= \sum_{q=0}^\infty \tilde\beta(q) n(q) = 
u^\top\sum_{q=0}^\infty \tilde\theta(q) n(q).
\label{est_spade}
\end{align}
This estimator has a variance
$\var(\check\beta) = \nu(\tilde\beta^2) = {\rm CRB}^{(\rm SPADE)}$,
requires no knowledge of any unknown parameter, and still works even
if the object happens to consist of a finite number of point sources,
with no restriction on their separations.  If $\theta_0 = 1$, the
constrained CRB can be derived in ways similar to Sec.~\ref{sec_ccrb}
and Appendix~\ref{sec_ccrb2}.

% The CRB given by Eq.~(\ref{CRB_spade}) and the efficient estimator
% given by Eqs.~(\ref{est_spade}) and (\ref{influ_spade}) are new
% results.  Reference~\cite{tsang17} was only able to approximate the
% CRB for the special case of Gaussian $\psi$, while the estimators
% proposed in Refs.~\cite{tsang17} and \cite{tsang18a} are either
% biased or restricted to the Gaussian-$\psi$ case and of unknown
% efficiency.

To estimate the odd moments of $F$ via SPADE, variations of the PAD
basis are needed \cite{tsang17,tsang18a}. The model is much more
complicated and a derivation of the CRB and the efficient estimator is
too tedious to work out here, but the upshot is the same: it can be
shown that the tangent space for the problem encompasses the whole
Hilbert space $\mathcal H$, the efficient influence can be retrieved
from the relation $\beta = \nu(\tilde\beta)$, and an unbiased and
efficient estimator is $\check\beta(n) = \int \tilde\beta(x) dn(x)$.

\subsection{Gaussian point-spread function}
More explicit results can be obtained and
the assumptions can be checked more carefully by assuming
the Gaussian point-spread function
\begin{align}
\psi(z) &= \frac{1}{(2\pi)^{1/4}}\exp\bk{-\frac{z^2}{4}}.
\label{psi_gauss}
\end{align}
The PAD basis becomes the Hermite-Gaussian basis, and it can be shown
that \cite{tnl,tsang17,yang16}
\begin{align}
H(q|y) &= \tau \exp\bk{-\frac{y^2}{4}} \frac{(y/2)^{2q}}{q!}.
\end{align}
The $C$ matrix in Eq.~(\ref{C_spade}) can be determined by expanding
$\exp(-y^2/4)=\sum_{j=0}^\infty (-y^2/4)^j/j!$, giving
\begin{align}
C_{qj} &= \iverson{j\ge q}\frac{\tau(-1)^{j-q}}{4^{j}q! (j-q)!}.
\end{align}
It is not difficult to check that the matrix inverse of $C$ is
\begin{align}
\tilde\theta_{2j}(q) &= (C^{-1})_{jq} = 
\iverson{q\ge j}\frac{4^{j}q!}{\tau(q-j)!},
\label{Cinv_spade}
\end{align}
which is a degree-$j$ polynomial of $q$.
$\sum_{q=0}^\infty \tilde\theta_{2j}(q) f(q)$ is the $j$th factorial
moment of $f$ and indeed equal to $\theta_{2j}$, since $H(q|y)$ is
Poisson and its factorial moment is
$\sum_{q=0}^\infty \tilde\theta_{2j}(q) H(q|y) = y^{2j}$
\cite{daley03}.  In general, each degree-$j$ moment of $H(q|y)$ is a
degree-$j$ polynomial of $y^2$, so each degree-$j$ moment of $f(q)$ is
a linear combination of the moments of $F$ up to degree $2j$.  All the
moments of $f$ are therefore finite as long as all the moments of $F$
are finite.  If $u$ has a finite number of nonzero entries, the
influence function given by Eqs.~(\ref{influ_spade}) is a polynomial
of $q$, so $\nu(\tilde\beta^2) < \infty$, and
$\tilde\beta \in \mathcal H$ is ensured.

\subsection{Bandlimited point-spread function}
Another important example is the bandlimited point-spread function
given by
\begin{align}
\Psi(k) &= \frac{\iverson{|k| < 1}}{\sqrt{2}}.
\label{rect}
\end{align}
$b$ is then the well known set of Legendre polynomials \cite{olver}.
Appendix~\ref{sec_legendre} shows the detailed calculations; here I
list the results only. Equation~(\ref{Hqy}) becomes
\begin{align}
H(q|y) &= \tau (2q+1)j_q^2(y),
\label{Hq_legendre}
\end{align}
where $j_q(y)$ is the spherical Bessel function of the first kind
\cite[Eq.~(10.47.3)]{olver}.  An influence function for estimating
$\theta_{2j}$ with $\theta_{2j} = \nu(\tilde\theta_{2j})$ is
\begin{align}
\tilde\theta_{2j}(q) &= 
 \iverson{q\ge j} \frac{(2j+1)!!(2j-1)!!}{\tau}\nchoosek{q+j}{2j},
\label{tilde_theta_spade}
\end{align}
where $!!$ denotes the double factorial \cite{olver}.
$\tilde\theta_{2j}(q)$ is a degree-$2j$ polynomial of $q$, so
$\tilde\beta(q)$ is also a polynomial of $q$ if $u$ contains a finite
number of nonzero entries.  As long as all the moments of $F$ are
finite, all the moments of $f$ can also be shown to be finite, and
$\nu(\tilde\beta^2) < \infty$ is ensured.

Notice that the direct-imaging theory in Sec.~\ref{sec_direct} breaks
down for this bandlimited point-spread function, as the second and
higher even moments of
$H(x) = \tau |\psi(x)|^2 = (\tau/\pi)\sinc^2(x)$ are all infinite. The
CRB in that case remains an open problem, although it is possible to
apodize the point-spread function optically such that all its moments
become finite and the semiparametric estimator given by
Eq.~(\ref{est_direct}) has a finite variance. For example, if
\begin{align}
\Psi(k) &\propto \iverson{|k|< 1}\exp\bk{-\frac{1}{k^2-1}},
\end{align}
then $\Psi(k)$ is infinitely differentiable despite the hard bandwidth
limit \cite{debnath} and all the moments of $|\psi(x)|^2$ are finite
\cite{tsang18a}.

\section{\label{sec_compare}Comparison of imaging methods}
The advantage of SPADE over direct imaging occurs in the
subdiffraction regime, where the width $\Delta$ of the object
distribution $F$ with respect to the origin is much smaller than the
width of the point-spread function $\psi$
\cite{tsang17,tsang18a,tsang19,tsang19a}.  As the width of $\psi$ is
normalized as $1$, the regime is defined as
\begin{align}
\Delta &\ll 1,
\end{align}
and the object moments scale as
\begin{align}
\theta_j &= \theta_0 O(\Delta^j).
\end{align}
With the attainable CRBs given by Eqs.~(\ref{CRB_direct}) and
(\ref{CRB_spade}) at hand, I can now compare direct imaging and SPADE
on the same semiparametric footing.  Consider the estimation of a
specific moment $\theta_k$ with
\begin{align}
u_j = \delta_{jk}.
\end{align}
For direct imaging in the subdiffraction regime, the image becomes
close to the point-spread function, viz.,
\begin{align}
f(x) &\approx \theta_0 H(x) =  N |\psi(x)|^2,
\end{align}
where $N$, the expected photon number received in total, is given by
Eq.~(\ref{N}). With $C_{jk} = \tau O(1)$ and
$\nu(\tilde\phi\tilde\phi^\top) = N O(1)$, the CRB becomes
\begin{align}
{\rm CRB}^{(\rm direct)} &= \frac{\theta_0^2}{N}O(1).
\label{CRB_direct2}
\end{align}
For SPADE on the other hand, notice that the $C$ and $C^{-1}$ matrices
are upper-triangular, meaning that
\begin{align}
f(q) &= N O(\Delta^{2q}),
\end{align}
and the CRB for estimating $\theta_{k}$, where $k$ is even, becomes
\begin{align}
{\rm CRB}^{(\rm SPADE)}
&= \frac{\theta_0^2}{N}O(\Delta^{k}),
\label{CRB_spade2}
\end{align}
which is much lower than Eq.~(\ref{CRB_direct2}) when $\Delta \ll 1$
and $k \ge 2$. This is consistent with earlier approximate results
\cite{tsang17,tsang18a}. An intuitive explanation of the enhancement,
as well as a discussion of the limitations of SPADE, can be found in
Ref.~\cite{tsang19a}. The constrained CRB with $\theta_0 = 1$ becomes
$[O(\Delta^k)-\theta_k^2]/N = O(\Delta^k)/N$, which is on the same
order of magnitude as the fundamental quantum limit \cite{tsang19}.

\begin{figure}[htbp!]
\centerline{\includegraphics[width=0.48\textwidth]{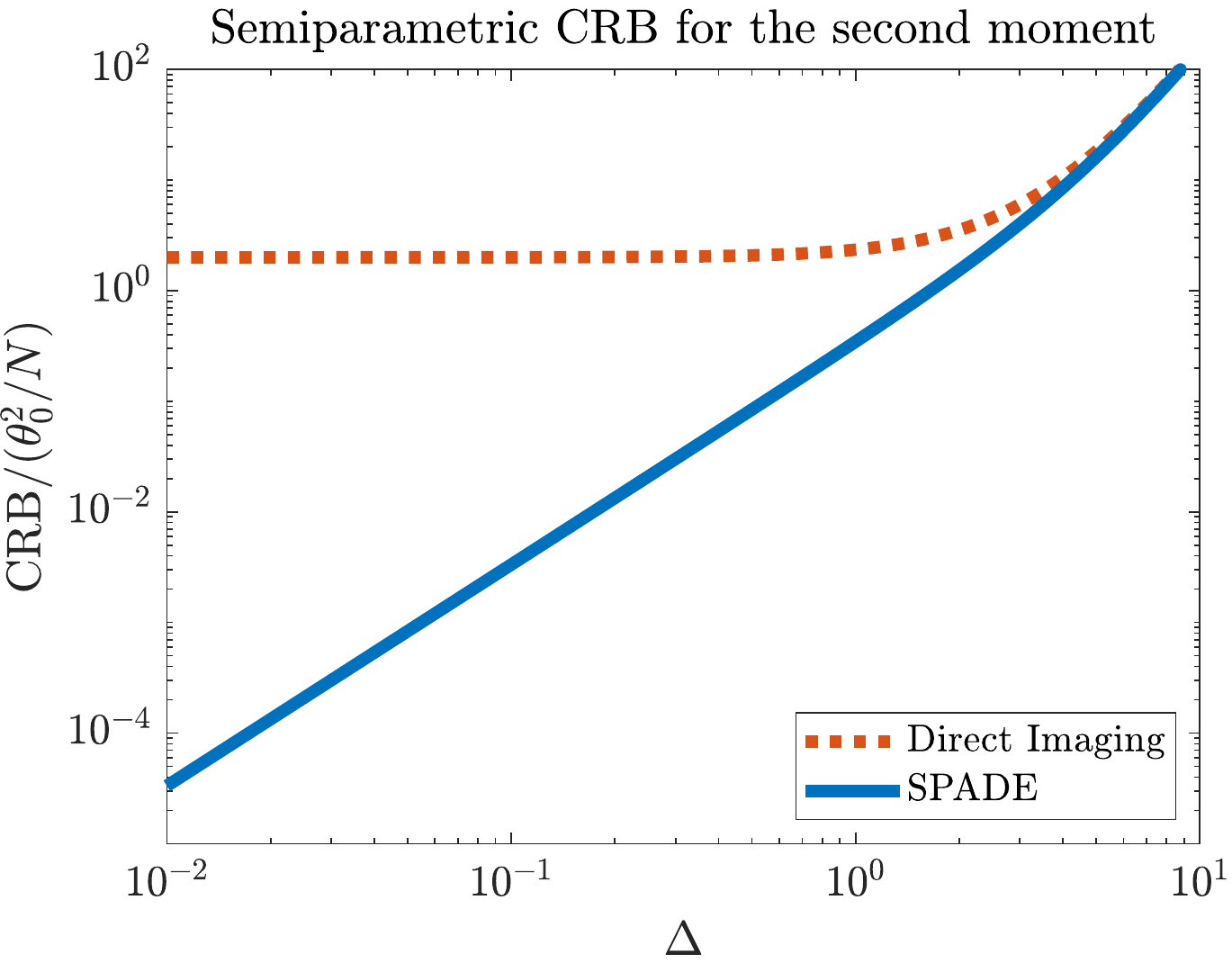}}
\caption{\label{semiparametric_CRB}The semiparametric CRBs for the
  second moment $\theta_2$ given by Eqs.~(\ref{CRB_direct_2nd}) and
  (\ref{CRB_spade_2nd}) versus the object size $\Delta$ in log-log
  scale, if the point-spread function is Gaussian and the object
  happens to be a flat top. Both the CRBs and $\Delta$ are normalized
  such that they are dimensionless.}
\end{figure}

More exact and pleasing results can be obtained if $\psi$ is Gaussian
and given by Eq.~(\ref{psi_gauss}).  Consider for example the
estimation of the second moment $\theta_2$.  For direct imaging, it
can be shown that
\begin{align}
{\rm CRB}^{(\rm direct)}
&= 
\frac{1}{\tau}\bk{2\theta_0 + 4\theta_2 + \theta_4} = 
\frac{\theta_0^2}{N} O(1).
\label{CRB_direct_2nd}
\end{align}
For SPADE on the other hand,
\begin{align}
%\tilde\beta(x) &= \frac{4}{\tau} q,
%\\
{\rm CRB}^{(\rm SPADE)} &= 
\frac{1}{\tau}\bk{4\theta_2 + \theta_4} 
= \frac{\theta_0^2}{N}O(\Delta^{2}),
\label{CRB_spade_2nd}
\end{align}
which not only beats direct imaging by a significant amount in the
subdiffraction regime but is in fact superior for all parameter
values. To further illustrate the difference between the two methods,
suppose that the object happens to be a flat top given by
\begin{align}
dF(y) &= \frac{\theta_0}{\Delta} \iverson{|y| < \Delta/2} dy.
\end{align}
Do note that the semiparametric CRBs do not assume the knowledge of
this object shape, which is specified here only for the purpose of
plotting examples of the CRBs. With $\theta_2 = \theta_0\Delta^2/12$
and $\theta_4 = \theta_0 \Delta^4/80$, Fig.~\ref{semiparametric_CRB}
plots Eqs.~(\ref{CRB_direct_2nd}) and (\ref{CRB_spade_2nd}) against
$\Delta$ in log-log scale.  The gap between the two curves in the
$\Delta\ll 1$ regime is striking.

With the constraint $\theta_0 = 1$, the CRBs become
\begin{align}
{\rm CRB}_{\theta_0=1}^{(\rm direct)}
&= 
\frac{1}{N}\bk{2 + 4\theta_2 + \theta_4-\theta_2^2} = 
\frac{O(1)}{N},
\\
{\rm CRB}_{\theta_0=1}^{(\rm SPADE)} &= 
\frac{1}{N}\bk{4\theta_2 + \theta_4-\theta_2^2} 
= \frac{O(\Delta^{2})}{N}.
\label{CCRB_spade2}
\end{align}
It is noteworthy that Eq.~(\ref{CCRB_spade2}) is exactly equal to the
quantum limit given by \cite[Eq.~(E15)]{tsang19a}, meaning that SPADE
is exactly quantum-optimal---at all parameter values---for estimating
the second moment. This is consistent with previous results concerning
the estimation of two-point separation \cite{tnl} and object size
\cite{tsang17,dutton19}, but note that the previous results assume
that the object shape is known, whereas the CRBs and the estimators
here assume the opposite.

\section{\label{sec_conc}Conclusion}
The semiparametric theory set forth solves an important and difficult
problem in incoherent optical imaging: the evaluation of the CRB and
the efficient estimation of object parameters when little prior
information about the object is available. The theory gives exact and
achievable semiparametric CRBs for both direct imaging and SPADE,
establishing the superiority and versatility of SPADE beyond the
special parametric scenarios considered by previous studies.

Despite the elegant results, the theory has a few shortcomings. On the
mathematical side, the conditions for the theory to hold seem
difficult to check in the case of direct imaging with a non-Gaussian
point-spread function, especially when the point-spread function has
infinite moments.  It is an open question whether this is merely a
technicality or a hint at a whole new regime of statistics. On the
practical side, the theory may be accused of assuming ideal conditions
for both measurements, such as infinitesimal pixels for direct
imaging, the availability of infinitely many modes for SPADE, perfect
specification and knowledge of the optical systems, and the absence of
excess noise. Reality is necessarily uglier, but the results here
remain relevant by serving as fundamental limits (via the
data-processing inequality \cite{ibragimov81,tsang19a}) and offering
insights into the essential physics. The theoretical and experimental
progress on SPADE and related methods so far
\cite{tnl,tsang17,tsang18a,dutton19,tsang19,zhou19,sliver,tnl2,nair_tsang16,lupo,tsang18,ant,lu18,rehacek17,yang17,kerviche17,chrostowski17,rehacek17a,rehacek18,backlund18,napoli19,yu18,prasad19,larson18,tsang_comment19,larson19,grace19,bonsma19,tsang19a,tang16,tham17,paur16,yang16,donohue18,parniak18,hassett18,zhou19a,parniak18,paur18,paur19}
has provided encouragement that the theory has realistic potential,
and the general results here should motivate further research into the
wider applications of quantum-inspired imaging methods.

An interesting future direction is to generalize the semiparametric
formalism for quantum estimation \cite{helstrom,holevo11}. By treating
the symmetric logarithmic derivatives of the quantum state $\rho$ as
the scores in the $\mathcal L_h^2(\rho)$ space proposed by Holevo
\cite{holevo11} and adopting a geometric picture \cite{fujiwara05}, a
quantum generalization of the semiparametric CRB can be envisioned,
but whether it can solve any important problem, such as the quantum
limit to incoherent imaging \cite{tsang19,zhou19}, remains to be seen.

\section*{Acknowledgments}
This work is supported by the Singapore National Research Foundation
under Project No.~QEP-P7.

\appendix
\section{\label{sec_proof}Proof of the semiparametric CRB
for Poisson processes}
Define the inner product between two random variables $\check r_1$ and
$\check r_2$ as
\begin{align}
(\check r_1,\check r_2) &= \expect\bk{\check r_1\check r_2},
\end{align}
and the norm as
\begin{align}
|||\check r||| &= \sqrt{(\check r,\check r)} = \sqrt{\expect(\check r^2)}.
\end{align}
Let the Hilbert space of zero-mean random variables be
\begin{align}
\check{\mathcal R} &= \BK{\check r: \expect(\check r) = 0,
\expect(\check r^2) < \infty},
\end{align}
and define
\begin{align}
\check{\mathcal T} &= \cspan(\check S) \subseteq \check{\mathcal R},
\end{align}
where $\check S$ is defined by Eq.~(\ref{ld}). Let
$\check\delta \in \check{\mathcal R}$ be any random variable that
satisfies
\begin{align}
\expect(\check\delta \check S) &= u.
\label{condition}
\end{align}
The semiparametric CRB is \cite{bickel93,tsiatis06}
\begin{align}
\expect(\check\delta^2) &\ge {\rm CRB} = \expect(\check\delta_{\rm eff}^2),
\label{CRB2}
\\
\check\delta_{\rm eff} &= \Pi(\check\delta|\check{\mathcal T}) = 
\argmin_{\check h \in \check{\mathcal T}} |||\check\delta - \check h|||.
\label{influ_eff2}
\end{align}
The proof can be done via a Pythagorean theorem \cite{tsiatis06}
without recourse to the Cauchy-Schwartz inequality or the existence of
$J^{-1}$. $\check S_j$ is called a score and $\check\delta$ an
influence in statistics \cite{bickel93,tsiatis06}; this paper uses the
same terminology for $S$ and $\tilde\beta$ in light of their
resemblance to the statistical quantities.

The resemblance can be turned into a precise correspondence for a
Poisson random measure by considering the subspace
$\check{\mathcal H} \subset \check{\mathcal R}$ of random variables
that are linear with respect to $n$. Any element
$\check h \in \check{\mathcal H}$ can be expressed as
\begin{align}
\check h &= Uh = \int h(x) \Bk{dn(x) - d\bar n(x)},
\end{align}
where $U$ is a surjective linear map from $\mathcal H$ to
$\check{\mathcal H}$.  Since
\begin{align}
(Uh_1,Uh_2) &= \Avg{h_1,h_2} \quad \forall h_1,h_2 \in \mathcal H
\end{align}
by virtue of Eq.~(\ref{poisson_covar}), $\check{\mathcal H}$ is
isomorphic to $\mathcal H$ and $U$ is unitary \cite{reed_simon}, and
since $\check{\mathcal T} \subseteq \check{\mathcal H}$ and
$\check S = US$, $\check{\mathcal T}$ is isomorphic to $\mathcal T$.
Picking a $\check\delta \in \check{\mathcal H}$ with
\begin{align}
\check\delta &= U\tilde\beta = 
\int \tilde\beta(x) \Bk{dn(x) - d\bar n(x)}
\end{align}
leads to 
\begin{align}
\expect(\check\delta\check S) &= \nu(\tilde\beta S) = u,
&
\check\delta_{\rm eff} &= U \tilde\beta_{\rm eff},
\end{align}
where $\tilde\beta_{\rm eff}$ is given by Eq.~(\ref{influ_eff})
because of Eq.~(\ref{influ_eff2}) and the isomorphisms. The CRB
becomes
\begin{align}
{\rm CRB} = \expect(\check\delta_{\rm eff}^2) = \nu(\tilde\beta_{\rm eff}^2),
\end{align}
which is Eq.~(\ref{CRB}).

\section{\label{sec_moments}The moment parameter space}
Define an $s\times s$ Hankel matrix with respect to a real-number
sequence $\theta = (\theta_0,\theta_1,\dots)^\top$ as
\begin{align}
M_{jk}^{(s)}(\theta) &= \theta_{j+k},
\quad
j,k \in \{0,1,\dots,s-1\}.
\label{hankel}
\end{align}
If $\theta$ is a moment sequence that arises from a nonnegative
measure $F$,
\begin{align}
w^\top M^{(s)} w = \int \bk{\sum_{j=0}^{s-1} w_j y^{j}}^2 dF(y) 
\label{represent}
\end{align}
is nonnegative for any real vector $w$, and all Hankel matrices are
positive-semidefinite, viz.,
\begin{align}
M^{(s)} &\ge 0
\quad
\forall s \in \mathbb N_1.
\label{Mpos}
\end{align}
Conversely, any sequence with Hankel matrices that obey
Eq.~(\ref{Mpos}) can be expressed in the form of Eq.~(\ref{moments})
with a nonnegative $F$ by virtue of Hamburger's theorem
\cite{schmudgen17}.

For the CRB to hold for a $p$-dimensional $\theta$, the parameter
space $\Theta$ should be an open subset of $\mathbb R^p$
\cite{ibragimov81,gorman}. Intuitively, the requirement makes sense
because all the parameters in $\theta$ are unknown and $\theta$ should
be allowed to vary in a neighborhood, otherwise the problem would be
overparametrized. If $\Theta$ is not an open subset, the parameter
space would be constrained and the CRB may be modified
\cite{gorman}. When all the moments are unknown parameters, consider
the set
\begin{align}
\Theta &= \BK{\theta: M^{(s)}(\theta) > 0\ \forall s \in \mathbb N_1}.
\label{Theta}
\end{align}
Each $\theta \in \Theta$ corresponds to a measure with infinite
support $r = \infty$ \cite{schmudgen17}. The proof can be done by
observing that the polynomial in Eq.~(\ref{represent}) has at most
$s-1$ zeros and the integral is strictly positive for any $w \neq 0$
if and only if $r \ge s$, and therefore the constraint for $\Theta$ is
satisfied if and only if $r = \infty$. For $r < \infty$, $F$ can be
expressed in terms of its support
$\{y_l: 0\le l \le r-1, y_l < y_{l+1}\}$ as
\begin{align}
dF(y) &= \sum_{l=0}^{r-1} F_l \iverson{y=y_l},
&
\frac{dF(y)}{dy} &= \sum_{l=0}^{r-1} F_l \delta(y-y_l).
\end{align}
In the context of optics, $r$ is the minimum number of point sources
that can describe the object distribution. The assumption of
Eq.~(\ref{Theta}) as the parameter space is consistent with the
infinite-support assumption for semiparametric estimation with mixture
models \cite[Sec.~6.5]{bickel93}, and it also makes intuitive sense,
at least as a necessary condition---with $r$ point sources, there are
only $2r$ unknown parameters, and the problem would be
overparametrized if all the moments are assumed to be unknown. Further
inequality constraints on $\theta$ may be needed to ensure the
convergence of the Taylor series in Eqs.~(\ref{parametric}) and
(\ref{f_spade}), although they should not affect the CRB as long as
$\theta$ obeys and stays away from them \cite{gorman}.

The boundary of $\Theta$ corresponds to measures with finite support
$r < \infty$.  If $s \le r$, then $M^{(s)}> 0$ and $M^{(s)}$ is
full-rank (rank $ = s$), but if $s > r$, I can write
\begin{align}
M^{(s)} &= V^\top \diag(F) V,
\\
V_{jk} &= y_l^k,
\quad
\diag(F)_{jk} = 
\iverson{0\le j\le r-1}F_j\delta_{jk}.
\end{align}
$V$ is the Vandermonde matrix and invertible since $\{y_l\}$ are
assumed to be distinct \cite{horn}.  With $M^{(s)} \ge 0$ and
$\diag(F) \ge 0$, Sylvester's law of inertia \cite{horn} implies that
the rank of $M^{(s)}$ is the same as the rank of $\diag(F)$, which is
$r$. In other words, the rank of $M^{(s)}$ is $\min(r,s)$, and any
finite $r$ means that $M^{(s)}$ is rank-deficient and does not satisfy
the strict $M^{(s)} > 0$ for all $s > r$. Whether the CRB still holds
for $\theta$ on the boundary is a difficult question, considering that
the parameter space here is infinite-dimensional and it is not obvious
how existing finite-dimensional results regarding the CRB on a
boundary \cite{gorman} can be applied.

\section{\label{sec_series}Series expansion of the object
  distribution}
Assume that the object measure $F$ can be expressed as the
orthogonal series 
\begin{align}
dF(y) &= \sum_{j=0}^\infty \xi_j g_j(y) dF^{(0)}(y)
\label{fourier}
\end{align}
with respect to a reference measure $F^{(0)}$, where
$\{g_j = \sum_{k=0}^\infty G_{jk} y^k: j\in\mathbb N_0\}$ are the
orthogonal polynomials defined by
\begin{align}
g_j(y) &= \sum_{k=0}^\infty G_{jk} y^k,
&
\int g_j(y) g_k(y) dF^{(0)}(y) &= \delta_{jk},
\end{align}
and $G$ is a lower-triangular matrix with nonzero diagonal entries
that can be obtained by the Gram-Schmidt procedure. Thus each
``Fourier'' coefficient $\xi_j$ can be expressed in terms of the
moment parameters as
\begin{align}
\xi_j &= \int g_j(y) dF(y) = \sum_{k=0}^\infty G_{jk}\theta_k,
\end{align}
which can be written as
\begin{align}
\xi &= G \theta, & \theta &= G^{-1}\xi.
\label{xi_theta}
\end{align}
Hence each $\theta$ corresponds to a set of coefficients $\xi$ that
can be used to represent $F$ via Eq.~(\ref{fourier}). It is
straightforward to transform the CRBs and the efficient estimators
derived in this paper for $\theta$ to those for $\xi$ via
Eqs.~(\ref{xi_theta}).

\section{\label{sec_tangent}Tangent space for the direct-imaging
  model}
Consider the tangent space $\mathcal T$ given by Eq.~(\ref{tangent2})
and the score functions given by Eq.~(\ref{S_direct}) for direct
imaging. First note that $S \subset \mathcal H$, as the Fisher
information $J_{jj} = \avg{S_j,S_j} = \nu(S_j^2)$ is assumed to be
finite for all $j$. Recent calculations in quantum estimation theory
suggest that this assumption is reasonable for any measurement
\cite{tsang19}. To prove $\mathcal T = \cspan(S) = \mathcal H$, the
standard method is to show that the only element in $\mathcal H$
orthogonal to $S$ is $0$ \cite{reed_simon}, that is,
\begin{align}
\avg{h,S_j} = 0\quad \forall j \in \mathbb N_0
\label{ortho}
\end{align}
implies $h = 0$ (almost everywhere with respect to $\bar n$).  Here I
list a few approaches with various levels of rigor.

The first approach is to consider the set of orthogonal polynomials
\begin{align}
a &= \BK{a_j(x) = A \tilde\phi(x): j \in \mathbb N_0,
\Avg{a_j,a_k} = \delta_{jk}},
\label{poly}
\end{align}
where $A$ is a lower-triangular matrix with nonzero diagonal entries
and can be determined by applying the Gram-Schmidt procedure to the
monomials $\tilde\phi(x)$ \cite{dunkl}. Under rather general
conditions on $f$, the polynomials form an orthonormal basis of
$\mathcal H$ \cite{dunkl}, viz.,
\begin{align}
\mathcal H &= \cspan(a),
\end{align}
and I can write Eq.~(\ref{ortho}) as
\begin{align}
\Avg{h,S_j} &= \sum_{k=0}^\infty \Avg{h,a_k}\Avg{a_k,S_j} = 0
\quad
\forall j \in \mathbb N_0,
\label{Sk}
\end{align}
or, more compactly,
\begin{align}
B^\top w &= 0, & w_k &= \avg{h,a_k}, &
B_{kj} &= \avg{a_k,S_j}.
\label{BTw}
\end{align}
If the only solution to Eq.~(\ref{BTw}) is $w = 0$, then the only
solution to Eq.~(\ref{Sk}) is $h = 0$. This is equivalent to the
condition that $B^\top$ is injective.

Integration by parts yields
\begin{align}
B_{kj} &= \frac{(-1)^j}{j!}\int  
a_k(x) \parti{^jH(x)}{x^j}dx
= \sum_{l=0}^\infty A_{kl} C_{lj},
\label{B_direct}
\end{align}
where $C$ is the same as Eq.~(\ref{C_direct}). Since both $A$ and $C$
are lower-triangular with nonzero diagonal entries, $B = AC$ is also
lower-triangular with nonzero diagonal entries, and $B^\top$ has a
well defined matrix inverse
$(B^\top)^{-1} = (B^{-1})^\top = A^{-\top}C^{-\top}$ in the sense that
\begin{align}
B^\top (B^\top)^{-1} = (B^\top)^{-1} B^\top = I,
\end{align}
where $I$ is the identity matrix; see Appendix~\ref{sec_triangular}
for details. If the matrices were finite-dimensional, the existence of
a matrix inverse would imply
\begin{align}
(B^\top)^{-1}(B^\top w) = [(B^\top)^{-1} B^\top] w = w,
\label{associative}
\end{align}
and the only solution to Eq.~(\ref{BTw}) would be $w = 0$. This proof
is not entirely satisfactory however, as Eq.~(\ref{associative})
assumes that the product of the infinite-dimensional matrices is
associative.  Associativity assumes that the order of the sums
involved in the matrix product can be interchanged, but it cannot be
guaranteed for infinite-dimensional matrices. In other words, the
existence of a matrix inverse for $B^\top$ may not imply that $B^\top$
is injective.

A more rigorous approach is to define
\begin{align}
\chi_y(x) &=
\sum_{j=0}^\infty y^j S_j(x),\quad y \in \mathcal Y \subset\mathbb R,
\end{align}
and notice that Eq.~(\ref{ortho}) implies
\begin{align}
\Avg{h,\chi_y} &= \int h(x) 
\sum_{j=0}^\infty y^j\frac{(-1)^j}{j!}\parti{^jH(x)}{x^j} dx
\\
&= \int h(x) H(x-y)dx = 0 \quad
\forall y \in \mathcal Y.
\label{complete}
\end{align}
The unique solution to Eq.~(\ref{complete}) is $h = 0$ if the family
$\{H(x-y): y \in \mathcal Y\}$ satisfies a property called
completeness in statistics \cite{lehmann98}.  For example, if $H$ is
Gaussian, $\{H\}$ is a full-rank exponential family for any open
subset $\mathcal Y \subset \mathbb R$ and therefore complete
\cite{lehmann98}. An even more rigorous formulation of this approach
\cite{bickel93} is to treat $\avg{h,\chi_y}$ as an operator that maps
$h \in \mathcal H$ to a function of $y$ in another Hilbert space, and
then show that the null space of the operator consists of only
$h = 0$. The proof again boils down to the requirement that $\{H\}$
should be complete; see Ref.~\cite[Sec.~6.5]{bickel93}.

\section{\label{sec_triangular}Inverse of an infinite-dimensional triangular matrix}
Let $C$ be an infinite-dimensional matrix with entries indexed by
$(j,k) \in \mathbb N_0^2$. Define its formal matrix inverse $C^{-1}$
as another infinite-dimensional matrix that satisfies
\begin{align}
\sum_{l=0}^\infty C_{jl}(C^{-1})_{lk} &= \delta_{jk}.
\end{align}
If $C$ is lower-triangular with nonzero diagonal entries,
viz.,
\begin{align}
C_{jk} &= 0 \textrm{ if }k > j,
&
C_{jj} &\neq 0,
\end{align}
then $C^{-1}$ can be found by a recursive relation as follows.  Define
$C^{(s)}$ as the $s\times s$ upper-left submatrix of $C$.  Write
$C^{(s+1)}$ and $(C^{-1})^{(s+1)}$ as the partitions
\begin{align}
C^{(s+1)} &= \begin{pmatrix}C^{(s)} & 0\\ c^\top & C_{ss}\end{pmatrix},
\\
(C^{-1})^{(s+1)} &= \begin{pmatrix}
(C^{-1})^{(s)} & 0\\ d^\top & (C^{-1})_{ss}\end{pmatrix}.
\end{align}
Given $(C^{-1})^{(s)} = (C^{(s)})^{-1}$,
\begin{align}
d^\top &= - \frac{c^\top (C^{(s)})^{-1}}{C_{ss}},
&
(C^{-1})_{ss} &= \frac{1}{C_{ss}},
\end{align}
and the recursion starts from $(C^{-1})^{(1)} = (C^{(1)})^{-1}$ with
one element $(C^{-1})_{00} = 1/C_{00}$. The matrix inverse of an
infinite-dimensional upper-triangular matrix can be defined in a
similar way.

Although the product of infinite-dimensional matrices may not be
associative, it can still be proved by induction that $D (Cu) = (DC)u$
for any vector $u$ if $D$ and $C$ are lower-triangular, because
\begin{align}
D(Cu) &= \sum_{k=0}^\infty D_{jk}\sum_{l=0}^\infty C_{kl}u_l
= \sum_{k=0}^j D_{jk}\sum_{l=0}^k C_{kl}u_l
\end{align}
involves finite sums only. Thus it is safe to assume that
$C^{-1}(Cu) = (C^{-1} C)u = u$ and $C(C^{-1}u) = (C C^{-1}) u = u$ if
$C$ is lower-triangular with nonzero diagonal entries.

\section{\label{sec_direct2}An alternative derivation
of the Cram\'er-Rao bound for direct imaging}
Consider the problem described in Sec.~\ref{sec_direct}.
Since the polynomials given by Eq.~(\ref{poly}) are an orthonormal
basis, the information matrix for the moment parameters can be
expressed as
\begin{align}
J_{jk} = \Avg{S_j,S_k} = \sum_{l=0}^\infty \avg{S_j,a_l}\avg{a_l,S_k},
\label{Jbasis}
\end{align}
meaning that $J = B^\top B$, where $B = AC$ is given by
Eq.~(\ref{B_direct}).  Ignoring the complications of multiplying
infinite-dimensional matrices, the CRB becomes
\begin{align}
J^{-1} = B^{-1}B^{-\top} = C^{-1}A^{-1}A^{-\top}C^{-\top}.
\end{align}
To evaluate $A^{-1}A^{-\top}$, notice that the orthonormality of $a$
can be written as
\begin{align}
\Avg{a_j,a_k} &= \sum_{l=0}^\infty 
\sum_{m=0}^\infty A_{jl} \Avg{\tilde\phi_l,
\tilde\phi_m} A_{km} = \delta_{jk},
\end{align}
where $\tilde\phi$ is the monomials given by Eq.~(\ref{monomials}).
In other words,
\begin{align}
A \nu(\tilde\phi\tilde\phi^\top) A^\top &= I,
&
A^{-1}A^{-\top} &= \nu(\tilde\phi\tilde\phi^\top),
\end{align}
giving
\begin{align}
J^{-1} &= C^{-1}\nu(\tilde\phi\tilde\phi^\top)C^{-\top}.
\end{align}
This leads to Eq.~(\ref{CRB_direct}) for the parameter
$\beta = u^\top\theta$.

\section{\label{sec_ccrb2}Alternative approaches
to the constrained Cram\'er-Rao bound}
One way of deriving the constrained CRB if $\theta_0$ is known is to
consider the information matrix $\tilde J$ with respect to the
parameters $\vartheta = (\theta_1,\theta_2,\dots)^\top$ without
$\theta_0$.  Then $\theta = (\theta_0,\vartheta^\top)^\top$, and
$\tilde J$ can be written as the submatrix of $J$, or
\begin{align}
J &= \begin{pmatrix}J_{00} & j^\top \\ j & \tilde J\end{pmatrix},
\end{align}
where $j$ is a column vector. Ignore the complications of dealing with
infinite-dimensional matrices and partition $J^{-1}$ similarly as
\begin{align}
J^{-1} &= 
\begin{pmatrix}E_{00} & e^\top \\ e & \tilde E\end{pmatrix}.
\end{align}
Then it is straightforward to show that
\begin{align}
\tilde J^{-1} &= \tilde E - \frac{e e^\top}{E_{00}}.
\end{align}
Let $\tilde\vartheta = (\tilde\theta_1,\tilde\theta_2,\dots)^\top$,
and observe that $\tilde\theta_0 = 1/C_{00}$ from
Eqs.~(\ref{tilde_theta}), (\ref{C_direct}), and (\ref{phi}).  Then
Eq.~(\ref{CRB_direct}) implies that
\begin{align}
  \tilde E &= \nu(\tilde\vartheta\tilde\vartheta^\top),
  \\
  e &= \nu(\tilde\vartheta \tilde\theta_0) = 
\frac{\nu(\tilde\vartheta)}{C_{00}}
= \frac{\vartheta}{C_{00}},
  \\
  E_{00} &= \nu(\tilde\theta_0 \tilde\theta_0) = \frac{\nu(1)}{C_{00}^2}
           =\frac{\phi_0}{C_{00}^2}.
\end{align}
Hence
\begin{align}
\tilde J^{-1} &= \nu(\tilde\vartheta\tilde\vartheta^\top) - 
\frac{\vartheta\vartheta^\top}{\phi_0},
\end{align}
which implies Eq.~(\ref{CCRB}) if the parameter of interest
is defined as $\beta = u^\top\theta$ with $u_0 = 0$.

Yet another way of deriving the constrained CRB can be found in
Ref.~\cite{gorman}, which can be shown to lead to the same result
here.

\section{\label{sec_tangent2}Tangent space for the SPADE model}
The proof is similar to the first approach in
Appendix~\ref{sec_tangent}. Consider $\mathcal H = \cspan(a)$ in terms
of an obvious orthonormal basis
\begin{align}
a &=\BK{a_j(q)= \delta_{jq}/\sqrt{f(j)}: j \in \mathbb N_0}.
\label{a_spade}
\end{align}
Any $h\in\mathcal H$ orthogonal to the $S$ given by
Eq.~(\ref{S_spade}) obeys
\begin{align}
\Avg{h,S_{j}} &= \sum_{k=0}^\infty \Avg{h,a_k}\Avg{a_k,S_{j}}
 = 0\quad
\forall j \in \mathbb N_0,
\end{align}
which can be written as
\begin{align}
B^\top w &= 0, \quad w_k= \Avg{h,a_k},
\\
B_{jk} &= \Avg{a_j,S_{k}} =  \frac{C_{jk}}{\sqrt{f(j)}}.
\label{B_spade}
\end{align}
As $C$ is upper-triangular with nonzero diagonal entries and $f > 0$
is assumed, $B^\top$ is lower-triangular with nonzero diagonal
entries, and induction can be used to prove that the only solution to
$B^\top w = 0$ is $w = 0$, or equivalently $h = 0$. Hence
$\mathcal T = \cspan(S) = \mathcal H$. The proof is easier than the
one in Appendix~\ref{sec_tangent} because $B^\top$ here is
lower-triangular rather than upper-triangular.

An alternative proof, similar to the second approach in
Appendix~\ref{sec_tangent} and Ref.~\cite[Sec.~6.5]{bickel93} but less
fruitful in this case, is to define
\begin{align}
\chi_y(x) &= \sum_{j=0}^\infty y^{2j} S_{j}(x),
\quad
y \in \mathcal Y \subset \mathbb R,
\end{align}
consider
\begin{align}
\Avg{h,\chi_y} &= \sum_{q=0}^\infty h(q) H(q|y) = 0,
\end{align}
and use the completeness of $\{H(q|y): y \in \mathcal Y\}$ to prove
the unique solution $h = 0$. If $H$ is Poisson, for example, then
$\{H\}$ is a full-rank exponential family and therefore complete
\cite{lehmann98}.

\section{\label{sec_spade2}An alternative derivation of the
  Cram\'er-Rao bound for SPADE}
Consider the problem described in Sec.~\ref{sec_spade}.  With the
orthonormal basis given by Eq.~(\ref{a_spade}) and the $B$ matrix
given by Eq.~(\ref{B_spade}), the information matrix with respect to
the moment parameters can again be expressed as $J = B^\top B$
according to Eq.~(\ref{Jbasis}). With Eq.~(\ref{B_spade}), $B^{-1}$
becomes
\begin{align}
(B^{-1})_{jq} &= (C^{-1})_{jq}\sqrt{f(q)}.
\end{align}
Ignoring the complications of multiplying infinite-dimensional
matrices, the CRB $J^{-1} = B^{-1}B^{-\top}$ is
\begin{align}
(J^{-1})_{jk} &= 
\sum_{q=0}^\infty (C^{-1})_{jq}f(q)(C^{-1})_{kq}
=C^{-1}DC^{-\top},
\end{align}
where $D$ is given by Eq.~(\ref{D}), and the CRB for
$\beta = u^\top\theta$ coincides with Eq.~(\ref{CRB_spade}).

\section{\label{sec_legendre} Calculations concerning 
SPADE for a bandlimited point-spread function}
Consider the point-spread function given by Eq.~(\ref{rect}).
The standard Legendre polynomials are defined in terms of
\begin{align}
\frac{1}{2}\int_{-1}^1 P_q(k)P_p(k) dk &= \frac{1}{2q+1}\delta_{qp},
\end{align}
such that the orthonormal version is
\begin{align}
b_q(k) &= \sqrt{2q+1} P_q(k).
\end{align}
The Fourier transform of the polynomial is
\cite[Eq.~(18.17.19)]{olver}
\begin{align}
\frac{1}{2}\int_{-1}^1 b_q(k)\exp(iky) dk &= \sqrt{2q+1} j_q(y),
\label{fourier_legendre}
\end{align}
where $j_q(y)$ is the spherical Bessel function of the first kind
\cite[Eq.~(10.47.3)]{olver}. Then Eq.~(\ref{Hq_legendre}) follows
from Eq.~(\ref{Hqy}) and (\ref{fourier_legendre}).

Let
\begin{align}
\tilde H(q|y) = \frac{H(q|y)}{\tau} = (2q+1)j_q^2(y).
\end{align}
From Ref.~\cite[Eq.~(10.60.2)]{olver}, one can derive the useful
formula
\begin{align}
\sum_{q=0}^\infty \tilde H(q|y) P_q(k) &= \sinc w
= 
\begin{cases}
(\sin w)/w, & w \neq 0,\\
1, & w = 0,
\end{cases}
\\
w &= y\sqrt{2-2k}.
\end{align}
For example, since $P_q(1) = 1$, one can check that
$\sum_{q=0}^\infty \tilde H(q|y) = 1$ in accordance with
Eq.~(\ref{conserve}).  Using the facts
\begin{align}
\sinc w &= 
\frac{1}{2}\int_{-1}^1 \exp(iw z) dz
= 
\sum_{l=0}^\infty \frac{(-1)^l w^{2l}}{(2l+1)!},
\\
\diff{w}{k} &= -\frac{y}{w},
\quad
P_q^{(j)}(1) = \left.\diff{^j P_q(k)}{k^j}\right|_{k=1},
\end{align}
it can also be shown that
\begin{align}
\sum_{q=0}^\infty \tilde H(q|y) P_q^{(j)}(1) &= 
\left.\diff{^j \sinc w}{k^j}\right|_{k=1}
= \frac{y^{2j}}{(2j+1)!!},
\end{align}
which leads to
\begin{align}
\sum_{q=0}^\infty f(q)P_q^{(j)}(1) 
&= \frac{\tau\theta_{2j}}{(2j+1)!!}.
\end{align}
An influence function for estimating $\theta_{2j}$ is hence
\begin{align}
\tilde\theta_{2j}(q) &= \frac{(2j+1)!!}{\tau} P_q^{(j)}(1),
\label{influ_spade_rect}
\end{align}
which obeys $\theta_{2j} = \nu(\tilde\theta_{2j})$.
To derive an explicit expression for $P_q^{(j)}(1)$,
consider the Rodrigues formula \cite[Eq.~(14.7.13)]{olver}
\begin{align}
P_q(k) &= \frac{1}{2^qq!} \diff{^q}{k^q}(k^2-1)^q,
\end{align}
which leads to
\begin{align}
P_q(k) &= \sum_{l=0}^q 
\nchoosek{q}{l}
\nchoosek{q+l}{l}
\bk{\frac{k-1}{2}}^l,
\\
P_q^{(j)}(1) &=  \iverson{q\ge j}(2j-1)!!\nchoosek{q+j}{2j},
\label{Pqj1}
\end{align}
and Eq.~(\ref{tilde_theta_spade}) results.

To bound the moments of $\tilde H$ and $f$, consider a lower bound on
the binomial coefficient for $j\ge 1$ given by \cite[pp.~1186]{cormen}
\begin{align}
\nchoosek{q+j}{2j} &\ge \frac{(q+j)^{2j}}{(2j)^{2j}}
\ge \frac{q^{2j}}{(2j)^{2j}},
\end{align}
such that each even moment of $\tilde H$ can be bounded as
\begin{align}
&\quad \sum_{q=0}^\infty \tilde H(q|y) q^{2j}  
\nonumber\\
&=
\sum_{q=0}^{j-1}\tilde H(q|y) q^{2j} +
\sum_{q=j}^{\infty}\tilde H(q|y) q^{2j}
\\
&\le (j-1)^{2j} + \frac{(2j)^{2j}}{(2j-1)!!}
\sum_{q=0}^\infty \tilde H(q|y) P_q^{(j)}(1)
\\
&= (j-1)^{2j}  + \frac{(2j)^{2j}y^{2j}}{(2j-1)!!(2j+1)!!}.
\end{align}
This means that each even moment of $f(q)$ is bounded as
\begin{align}
\nu(q^{2j}) 
& \le \tau
\Bk{(j-1)^{2j}\theta_0 + 
\frac{(2j)^{2j}\theta_{2j}}{(2j-1)!!(2j+1)!!}}.
\end{align}
With $\nu(q^0) = \nu(1) = \tau\theta_0$, $\nu(q^{2j})<\infty$ for all
$j \in \mathbb N_0$ as long as $\theta_0$ and $\theta_{2j}$ are
finite. Odd moments can be bounded via the Cauchy-Schwartz inequality
$[\nu(q^j)]^2 \le \nu(1)\nu(q^{2j})$. Hence all the moments of $f$ are
finite as long as all the moments of $F$ are finite.

%\section{\label{sec_odd}Odd-moment estimation with SPADE}

\bibliography{research2}

\end{document}